\documentclass[pra,twocolumn,groupedaddress,shownopacs,amssymb,floatfix]{revtex4-2}
\usepackage{graphicx,psfrag,amsmath,amssymb}
\usepackage[usenames, dvipsnames]{color}
\usepackage{braket}
\usepackage{amsmath}
\usepackage{amssymb}
\usepackage{gensymb}
\usepackage{makeidx}
\usepackage{mathtools}
 
\usepackage[normalem]{ulem}
\usepackage{float}
\usepackage[T1]{fontenc} 
\usepackage{balance}
\usepackage{flushend}
\usepackage{lineno}
\usepackage{epsfig}
\usepackage{subfigure}
\usepackage{mathtools}
\usepackage{setspace}
\usepackage{bm}
\usepackage{ulem}
\usepackage{times}
\usepackage{enumitem}
\usepackage{xcolor}
\usepackage{multirow}
\usepackage{soul}
\usepackage{dcolumn}
\usepackage{balance}
\usepackage{siunitx}
\usepackage{epstopdf}
\usepackage{braket}
\usepackage{color}
\usepackage{diagbox}
\usepackage[utf8]{inputenc}
\usepackage{hyperref}
\hypersetup{
    colorlinks=true,
    linkcolor=Blue,
    citecolor=Blue,
    filecolor=Blue,
    urlcolor=Blue,
    }

\hypersetup{
    colorlinks=true,
    linkcolor=Blue,
    citecolor=Blue,
    filecolor=Blue,
    urlcolor=Blue,
    }
\begin{document}
\title{Structural transition and fragmentation of vortex lattices in rotating tilted dipolar \\Bose-Einstein condensate}
\author{Mamta Ale}
\thanks{These authors contributed equally to this work.}
\affiliation{Department of Physics, Central University of Rajasthan, Ajmer - 305817, India}
\author{Harsimranjit Kaur}
\thanks{These authors contributed equally to this work.}
\affiliation{Department of Physics, Central University of Rajasthan, Ajmer - 305817, India}
\author{Kuldeep Suthar}
\affiliation{Department of Physics, Central University of Rajasthan, Ajmer - 305817, India}

\date{\today}
\begin{abstract}
 We investigate the vortex lattices of harmonically confined quasi-two-dimensional tilted rotational dipolar Bose-Einstein condensates. By employing an extended Gr\"oss-Pitaevskii equation for a rotating condensate, we reveal the structural transformation of vortices from square to triangular lattices as the tilt of dipolar bosons relative to the polarization axis approaches a critical angle. When the tilt of the magnetic dipoles surpasses the magic angle, the condensate elongates diagonally and becomes devoid of vortices. Moreover, we include the Lee-Huang-Yang correction, which enables the formation of vortices in the elongated condensate. Additionally, when dipoles are oriented perpendicular to the polarization axis, the Lee-Huang-Yang correction results in the fragmentation of condensates under strong rotation. The quench dynamics of the rotational frequency demonstrate the development of vortex lattices; however, with a strong rotational quench, the condensate remains free of vortices. Our numerical analysis highlights the beyond mean-field effects of the rotational properties of anisotropic dipolar bosons, which can be observed in current dipolar quantum gas experiments.
 
\end{abstract}
\maketitle
\section{Introduction}\label{Introduction}
Bose–Einstein condensates (BECs) characterized by long-range and anisotropic dipole–dipole interactions (DDI) present a distinctive opportunity for the engineering of novel quantum many-body phenomena within the realm of ultracold atomic physics. It is facilitated by pioneering experiments utilizing atoms with large magnetic moments, including
$^{52}\mathrm{Cr}$~\cite{griesmaier_05,Lahaye_07}, $^{164}\mathrm{Dy}$~\cite{lu_11,Tang_18}, and $^{168}\mathrm{Er}$~\cite{aikawa_12,Trautmann_18,Bigagli_24}. The $s$-wave contact interaction is localized and spherically symmetric, while the dipolar interaction is nonlocal, anisotropic, and affects all partial waves. The interplay between isotropic short-range contact interactions and anisotropic long-range dipolar interactions gives rise to rich physics in dipolar Bose–Einstein condensates (dBECs), including anisotropic superfluidity~\cite{lahaye_09}, roton-like excitations and instability~\cite{santos_03,Chomaz_18}, and self-bound quantum droplets stabilized by beyond-mean-field fluctuations~\cite{petrov_15,Barbut_16}. Rotation of the condensate leads to the formation of quantized topological excitations, which are referred to as vortices and demonstrates the superfluid nature of quantum systems~\cite{abad_09}. The fast-rotating condensates break down into two fragments in anisotropic potentials~\cite{kumar_16,brito_20,dutta_23}. 
Numerous theoretical studies~\cite{cooper_05,zhang_05,lahaye_09,Lahaye_07,Baranov_08,santos_03,kumar_12,chomaz_16,wilson_10} and recent dipolar quantum gas experiments~\cite{klaus_22,bland_23,
casotti_24} have explored the rotational properties of dBECs. The anisotropic character of the DDI can be controlled by tilting the orientation of the dipoles relative to the confinement geometry, and thus the effective interaction can be tuned from repulsive to attractive and can even be nullified~\cite{Edmonds_20,Tang_18,chomaz_19}. This tunability enables access to diverse stability regimes and collective structures, particularly in a quasi-two-dimensional (quasi-2D) trap where the condensate becomes highly sensitive to dipolar anisotropy~\cite{martin_12,raghunandan_15,chen_17,sabari_25}. The rotational properties of dBEC are strongly affected by the anisotropic character of the dipolar interaction and its relative strength with respect to contact interactions~\cite{yi_06,komineas_07,abad_09,binjen_09,cai_18}. The critical rotational frequency of vortex nucleation depends on the strength of the dipolar interaction as well as the orientation of the dipoles~\cite{fetter_74,malet_11,martin_17}. The nontrivial lattice structures that include half-quantum vortex and fractional skyrmion states of rotating binary dBECs have also been studied~\cite{zhao_13,shirley_14,zhang_16,kumar_17,dong_17}. 
 
Beyond the description of the mean-field, the Lee-Huang-Yang (LHY) correction accounts for the effects of quantum fluctuations arising from zero-point motion of Bogoliubov excitations \cite{petrov_15,li_18}. The LHY term introduces a repulsive beyond-mean-field nonlinearity that becomes crucial in regimes where the mean-field interactions are suppressed or nearly canceled. In particular, when competing interactions lead to a strong reduction of the resultant mean-field contribution, the LHY correction may dominate the effective nonlinear response of the condensate, stabilizing otherwise unstable quantum states. This mechanism has enabled the realization of self-bound quantum droplets in binary condensates~\cite{cabrera_18,cheiney_18,ferioli_19,yang_23}. In addition, the interplay between LHY-induced quantum fluctuations and nonlocal interactions plays a central role in the emergence of novel quantum phases such as supersolids~\cite{norcia_21,bland_22} and quantum droplets~\cite{tanzi_19,bottcher_19,chomaz_19,wachtler_16, edler_17,wachtler_16}. Recently, the nucleation dynamics and pinning of vortices in novel supersolid phases have been examined~\cite{gallemi_20,ancilotto_21}. The stability of few quantum vortices in dBEC with LHY correction has recently been examined~\cite{Sabari_26}. Despite recent theoretical studies and experimental progress, the role of anisotropic tilt on the rotational properties has not been explored yet.

In the present work, we examine the vortex lattices of rotating tilted dBEC confined in a quasi-two-dimensional harmonic trap. Here, we employ an extended Gr\"oss-Pitaevskii formalism with rotation and LHY correction to study the anisotropic dipolar bosons. We reveal the structural transition of vortex lattices as the orientation of the dipoles is varied relative to the polarization axis. Under fast rotation, when dipoles are polarized perpendicular to the rotation axis (strongly attractive DDI), the condensate acquires an elliptical shape and remains devoid of vortices. However, in the presence of LHY correction, the quantum fluctuations stabilize the vortices. For a large number of atoms, the density of the condensate breaks down into two fragments. Finally, we present the rotational quench dynamics of dBEC, where the sudden quench of rotational frequency leads to the vortex nucleation, while at strong rotation the system is dynamically trapped in a vortex-free state in the absence of surface mode instabilities.  

The remainder of the paper is organized as follows. In Sec.~\ref{THEORY AND METHODS}, we present the formulation of the extended Gr\"oss-Pitaevskii equation for a rotating quasi-two-dimensional dipolar Bose–Einstein condensate, including DDI and LHY effects. In Sec.~\ref{Numerical_Results}, we discuss the formation and evolution of vortex lattices under varying dipole tilt and rotational frequency, along with the role of LHY corrections and dynamical nucleation of vortices. Finally, we summarize our findings and outline future perspectives in Sec.~\ref{Conclusion}.
\section{THEORY AND METHODS}\label{THEORY AND METHODS}
\begin{figure}[h]
\includegraphics[width=\linewidth]{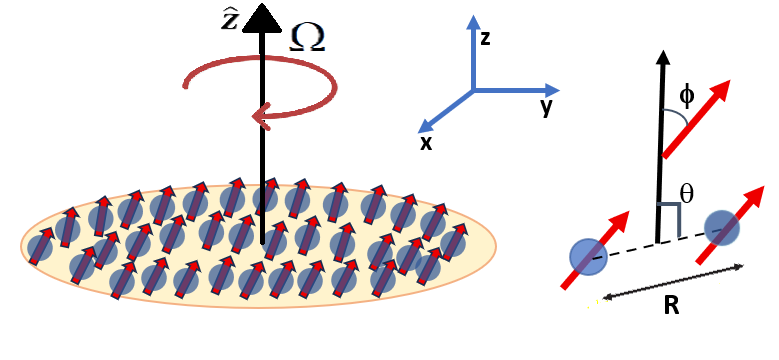}
  \caption{Schematic representation of the dipole-dipolar interaction in quasi-2D anisotropic dipolar Bose-Einstein condensate. The dipoles are polarized along the $z$-axis (left). The right panel of the figure shows the spatial and angular distributions, the interparticle displacement (separation) vector $\mathbf{R}$ and the polar (\( \theta \)) and azimuthal (\( \phi \)) angles. The angle between the separation vector and $z$-axis is $90^{\degree}$. The variation of the tilt angle $\phi$ between dipoles and polarization axis determines the strength and sign of the interaction, and thus change the character from repulsive to attractive. Here, $\Omega$ is the rotational frequency of the dipolar condensate.}
  \label{anisotropy}
\end{figure}

We consider a dipolar Bose-Einstein condensate of $N$ atoms confined in a quasi-2D pancake-shaped harmonic trap at zero temperature. The dynamics of an anisotropic rotating dBEC is governed by the extended Gr\"oss–Pitaevskii equation (eGPE) that includes the beyond mean-field LHY correction~\cite{kumar_17,wachtler_16a,baillie_18}
\begin{align}
i\hbar \frac{\partial \psi(\mathbf{r},t)}{\partial t} = &
-\frac{\hbar^2}{2m}\nabla^2 \psi(\mathbf{r},t) + {V_{\text{trap}}(\mathbf{r})} \psi(\mathbf{r},t) \nonumber \\
&+ U|\psi(\mathbf{r},t)|^2 \psi(\mathbf{r},t) \nonumber \\
&+ N \int V_{dd}(\mathbf{r} - \mathbf{r'})|\psi(\mathbf{r}',t)|^2 d\mathbf{r}' \,\psi(\mathbf{r},t) \nonumber  \\
&-\Omega L_z \psi(\mathbf{r},t)+ \frac{\eta_{\rm QF} \hbar^2 N^{3/2}}{m} |\psi(\mathbf{r},t)|^3 \psi(\mathbf{r},t).
\label{gpe_3d_equation}
\end{align}
Here, the harmonic trapping potential of the condensate is taken to be fully anisotropic and can be written as  
\begin{equation}
V_{\text{trap}}(\mathbf{r}) = \frac{1}{2}m\left( \omega_x^2 x^2 + \omega_y^2 y^2 + \omega_z^2 z^2 \right),
\end{equation}
where $\omega_x, \omega_y,$ and $\omega_z$ are the trap frequencies along the respective axis. The short-range contact interaction is $U={4\pi \hbar^2 a_{s} N}/{m}$ with $a_{s}$ and $m$ as the $s$-wave scattering length and atomic mass of the species, respectively. For dipolar atoms possessing magnetic moments, the interaction potential $V_{dd}$ is expressed as~\cite{kumar_15,lahaye_09,Sanjay_26}
\begin{equation}
  V_{dd}(\mathbf{R}) = \frac{\mu_0 \mu^{2}_{M}}{4\pi}\left(\frac{3\cos^2\phi-1}{2}\right) \left( \frac{1 - 3 \cos^2\theta}{|\mathbf{R}|^3}\right),
\end{equation}
where $\mathbf{R} = \mathbf{r} - \mathbf{r}'$ is the separation between the position of two dipoles and $\theta$ is the angle between the separation vector $\mathbf{R}$ and the polarization axis $z$, as shown in Fig.~\ref{anisotropy}. Here, $\mu_0$ is the vacuum permeability and $\mu_{M}$ is the magnetic dipole moment of a single atom. The angle of inclination of both dipole moments $\mu_{M}$ relative to the $z$-direction is $\phi$.

Our method accounts for a pancake-shaped two-dimensional confinement, where most of the dipolar atoms are concentrated near the transverse plane. The trapping frequencies satisfy $\omega_{z}/\omega_{\rho} \gg 1$, with $\rho = (x,y)$ is radial coordinate and  $\omega_{\rho}$ is corresponding trapping frequency. With the quasi-2D condition, it may be assumed that the separation vector or location of dipoles \textbf{R} is near a plane perpendicular to the $z$-direction, i.e. $\theta \approx 90^{\degree}$. The dBEC rotates about the $z$-axis with angular velocity $\Omega$ and $L_{z}$ is the $z$-component of the angular momentum operator defined as $L_z\equiv -i\hbar(x\partial_y-y\partial_x)$. The last term of eGPE [Eq.~\eqref{gpe_3d_equation}] is the LHY correction term that approximates the lowest-order quantum fluctuation contributions to the energy functional which appear to be repulsive in two- and three-dimensional systems~\cite{wachtler_16,wachtler_16a,li_18}.

The scaled form of eGPE is obtained by introducing a reference frequency $\overline{\omega}$ and the corresponding oscillator length $l = \sqrt{\hbar/(m\overline{\omega})}$. Defining the reduced variables $\tilde{\mathbf{r}} = \mathbf{r}/l$, $\tilde{R} = R/l$, $\tilde{a_{s}} = a_{s}/l$, $\tilde{a_{dd}} = a_{dd}/l$, $\tilde{t} = \overline{\omega}t$, and $\tilde{\psi} = l^{3/2}\psi$. After removing the overhead tilde from all variables, the dimensionless eGPE can be written as 
\begin{align}
i \frac{\partial \psi(\mathbf{r},t)}{\partial t} = &
- \frac{1}{2}\nabla^2 \psi(\mathbf{r},t)+ \frac{1}{2}\left(\gamma^2 x^2 + \nu^2 y^2 + \lambda^2 z^2\right)\psi(\mathbf{r},t) \nonumber \\
&+ 4\pi N a |\psi(\mathbf{r},t)|^2 \psi(\mathbf{r},t) \nonumber \\
&+ 3N a_{dd} \int V_{dd}^{3D}(\mathbf{R}) |\psi(\mathbf{r}',t)|^2 d\mathbf{r}' \, \psi(\mathbf{r},t) \nonumber \\
&-\Omega L_z \psi(\mathbf{r},t)
+\eta_{QF}N^{3/2}|\psi(\mathbf{r},t)|^3 \psi(\mathbf{r},t),
\label{dimensionaless_3dgpe}
\end{align}
with the dipolar kernel in dimensionless units is $\tilde{V}_{dd}(\mathbf{R}) = (3\cos^2\phi-1)/2 \times (1 - 3 \cos^2 \theta)/|\mathbf{R}|^3$. Here, the aspect ratios of the trap are defined as $\gamma = \omega_x / \overline{\omega}$, $\nu = \omega_y / \overline{\omega}$, and $\lambda = \omega_z / \overline{\omega}$, where $\overline{\omega} = (\omega_x \omega_y \omega_z)^{1/3}$ is the geometric mean of the trapping frequencies. The coefficient $\eta_{\text{QF}}$ corresponds to the LHY correction, which incorporates the effect of quantum fluctuations beyond the mean-field approximation. It is given by~\cite{young_22}
\begin{equation}
\eta_{\rm QF} = \frac{128}{3}\sqrt{\pi a^5}\, Q_5(\epsilon_{dd}),
\end{equation}
where the auxiliary function
\begin{equation*}
  Q_5(\epsilon_{dd}) = \int_0^1 dx \, (1 -\epsilon_{dd}+ 3x^2\epsilon_{dd})^{5/2},
\end{equation*}
can be evaluated to produce an analytic expression for the coefficient of quantum fluctuations~\cite{Bisset_16}
\begin{equation}
  \eta_{\rm QF} \approx \frac{128}{3}\sqrt{\pi a^5}\left(1 + \frac{3}{2}\epsilon_{dd}^2\right),
\end{equation}
where $\epsilon_{dd}= a_{dd}/a_{s}$ is the ratio of the dipolar and $s$-wave scattering lengths. For non-dipolar atoms with $\epsilon_{dd} = 0$, the above coefficient recovers the LHY-limit of repulsive hard-sphere atoms~\cite{lee_57}. For stronger axial confinement, the condensate remains in the ground state along the axial direction. Thus, the $z$-component wave function can be well approximated by a Gaussian function $\phi(z) = 1/(\pi d_{z}^{2})^{1/4}\times\exp(-z^{2}/2d_{z}^{2})$ with $d_{z} = 1/\sqrt{\lambda}$ is the axial harmonic oscillator length. The wavefunction $\psi(\mathbf{r},t)$ can be written as 
\begin{align}
\psi(\mathbf{r},t)=\psi_{2\text{D}}(x,y,t)\phi(z).
\label{2d_wavefunction}
\end{align}
Here, $\psi_{2D}(x,y,t)$ represents the effective 2D wave function describing radial dynamics. Using the above ansatz in Eq.~(\ref{dimensionaless_3dgpe}) and further integrating over the $z$-direction, we obtain the effective 2D equation for a disk-shaped dBEC
\begin{align}
i \frac{\partial \psi_{2D}(\boldsymbol{\rho},t)}{\partial t} 
&= \Bigg[ 
-\frac{\nabla^2_{\rho}}{2} 
+ \frac{\gamma^2 x^2 + \nu^2 y^2}{2}  
 + \frac{4\pi a N |\psi_{2D}|^2}{\sqrt{2\pi} d_z}\nonumber \\ 
&+ 3 a_{dd} N 
\int \frac{d\mathbf{k}_\rho}{(2\pi)^2} e^{-i\mathbf{k}_\rho \cdot \boldsymbol{\rho}} 
\tilde{n}_\rho(\mathbf{k}_\rho,t) 
\tilde{V}^{(d)}_\rho(\mathbf{k}_\rho) \nonumber \\ 
&-\Omega L_z + \zeta|\psi_{2d}|^3\Bigg] \psi_{2D}(\boldsymbol{\rho},t),
\label{eq:GP}
\end{align}
where $k_\rho = \sqrt{k_x^2 + k_y^2}$, and $\tilde{n}_\rho(\mathbf{k}_\rho,t)$ and $\tilde{V}_\rho^{(d)}(\mathbf{k}_\rho)$ are the Fourier transforms of the two-dimensional density and dipolar potential, respectively. In momentum space, the density can be expressed as~\cite{Goral_02,kumar_15}
\begin{align*}
 \tilde{n}_\rho(\mathbf{k}_\rho,t) 
&= \int d\boldsymbol{\rho} \, e^{i\mathbf{k}_\rho \cdot \boldsymbol{\rho}} 
|\phi_{2D}(\boldsymbol{\rho},t)|^2.
\end{align*}
 The dipolar potential can be written as
\begin{align}
\tilde{V}^{(d)}_{\rho}(\mathbf{k}_\rho) 
&= \frac{1}{2\pi}\left(\frac{3 \cos^2\phi-1}{2}\right) \int_{-\infty}^\infty dk_z 
\left[ \frac{3k_z^2}{k^2} - 1 \right] |\tilde{n}_\parallel(k_z)|^2 \nonumber \\
&= \left(\frac{3 \cos^2\phi-1}{2\sqrt{2\pi} d_z}\right) \left[ 2 - 3\sqrt{\pi}\,\xi \, e^{\xi^2}\{1 - \text{erf}(\xi)\} \right],
\end{align}
where $\xi = \frac{k_\rho d_z}{\sqrt{2}}$ and \text{erf}$(x)$ is the complementary error function of $x$. It should be noted that the integrated dipolar 2D potential depends on the axial aspect ratio 
$\lambda$. The dipole-dipole interaction depends on the dipolar strength and angle of inclination $\phi$ as $g_d^{2d}\left(3\cos^2\phi-1\right)/2$. Here, the dimensionless 2D dipolar interaction strength depends on the number of atoms $N$, axial oscillator length $d_{z}$, and dipolar scattering length $a_{dd}$ as $g_{d}^{2d} = (3 a_{dd}N)/(\sqrt{2\pi}d_z)$.

The nucleation of vortices occurs at values $\Omega \geqslant \Omega_c$, where $\Omega_c$ is the critical angular velocity. The arrangement of vortices in a vortex lattice strongly influenced by the strength of the dipolar interaction, which can be controlled through the tilt of magnetic dipoles. In non-dipolar atomic systems, where only short-range interactions are present, vortices tend to form triangular Abrikosov lattices, as reported in previous studies~\cite{abo_shaeeer_01, abo_shaeeer_02, haljan_01, schweikhard_04,kumar_19}, while for dipolar systems, the presence of long-range dipole interactions can drive transitions from the triangular vortex lattice to less-symmetric square lattice~\cite{zhang_05}. The last term of Eq.~\eqref{eq:GP} corresponds to the LHY correction which is repulsive in nature. As the ground state of the condensate undergoes zero-point energy fluctuations due to Heisenberg’s uncertainty principle, the LHY interaction accounts for the quantum fluctuations. The 2D dimensionless form of the LHY correction is given by the parameter $\zeta$ as
\begin{align}
  \zeta=\sqrt\frac{2}{5}\frac{\eta_{QF}}{\pi^{3/4}d_z^{3/2}}=\frac{128 \sqrt{2}a^{5/2}N^{3/2}}{3 \sqrt{5} \pi^{1/4}d_z^{3/2}}\left(1 + \frac{3}{2}\epsilon_{dd}^2\right),
\end{align}
which balances the attractive DDI and thus stabilizes the quasi-2D dipolar condensates. 

The total energy including the LHY correction corresponding to Eq.~\eqref{eq:GP} can be written in terms of the radial vector $\rho$ as
\begin{align}
E &= \int d\boldsymbol{\rho} \Bigg[
\frac{1}{2} |\nabla_{\rho} \psi_{2D}|^2
+ \frac{1}{2} \left( \gamma^2 x^2 + \nu^2 y^2 \right) |\psi_{2D}|^2 \nonumber \\
&\quad 
- \Omega \, \psi_{2D}^* L_z \psi_{2D} + \frac{2\pi a N}{\sqrt{2\pi} d_z} |\psi_{2D}|^4
+ \frac{2}{5} \zeta |\psi_{2D}|^5 \nonumber \\
&\quad + \frac{1}{2} \int d\boldsymbol{\rho}' \,
|\psi_{2D}(\boldsymbol{\rho})|^2
V^{(d)}_{\rho}(\boldsymbol{\rho} - \boldsymbol{\rho}')
|\psi_{2D}(\boldsymbol{\rho}')|^2
\Bigg],
\label{eq:energy}
\end{align}
where, due to the circular symmetry of the condensate, the aspect ratios $\gamma$ and $\nu$ are taken to be one.

The DDI can be tuned for any specific value of $g_{d}^{2\mathrm{d}}$ with a fixed number of atoms by changing the inclination by a tilt angle $\phi$. Varying the tilt angle $\phi$ from $0^{\degree}$ to $90^{\degree}$ modifies the effective two-dimensional dipolar interaction strength from $g_d^{2\mathrm{D}}$ to $-g_d^{2\mathrm{D}}/2$, respectively, while it becomes zero at the magic angle $\phi_m=\cos^{-1}(1/\sqrt{3})\approx 54.73^\circ$. Thus, the DDI can be tuned continuously from repulsive to attractive by controlling the angle of inclination $\phi$ with respect to the polarization axis. This anisotropy-induced tunability of dipolar interaction is the central focus of the present work. We tune the angle $\phi$ to different values from $0^{\degree}$ to $90^{\degree}$ and observe various vortex-lattice geometries and dynamical phenomena of rotating condensate. At tilt angles $\phi_{m}<\phi\leqslant 90^{\degree}$, the vortex lattices get distort with attractive DDI, and for perpendicular orientations $\phi\approx90^{\degree}$, the condensate (with larger $N$)collapses due to highly attractive DDI; in such cases the beyond mean-field quantum fluctuation interaction in the form of the Lee-Huang-Yang correction term becomes significant. 

\section{Results and Discussions}\label{Numerical_Results}
We examine the structure of vortex lattices in rotating dipolar condensate and the role of the inclination of magnetic dipoles. To this end, we consider a strongly dipolar Bose-Einstein condensate $\mathrm{}{^{164}\mathrm{Dy}}$ with $s$-wave scattering length $a_s = 100~a_0$ and a dipolar scattering length $a_{dd} = 130~a_0$~\cite{Youn_10}. The characteristic harmonic oscillator length is set as $l=1$\textmu m. Primarily, we assume the number of particles $N=1000$; however, we reveal the fragmentation of the ground state of dBEC with vortex lattices for a larger number of atoms. The latter occurs for the fast rotating condensate (i.e. $\Omega$ is close to the radial trapping frequency) of dBEC in which the perpendicular orientation of the dipoles is stabilized by the LHY correction. We first discuss rotating condensate below the magic angle $\phi_{m}$ (repulsive DDI), and then further above $\phi_{m}$ (attractive DDI) in the presence of LHY correction. To understand the dynamical properties, we finally present the response to the quench of the rotational frequency for different orientations of dipoles.   

\subsection{Effects of anisotropy of dipoles} 
\label{section_a}
\begin{figure}[h]
 \includegraphics[width=\linewidth]{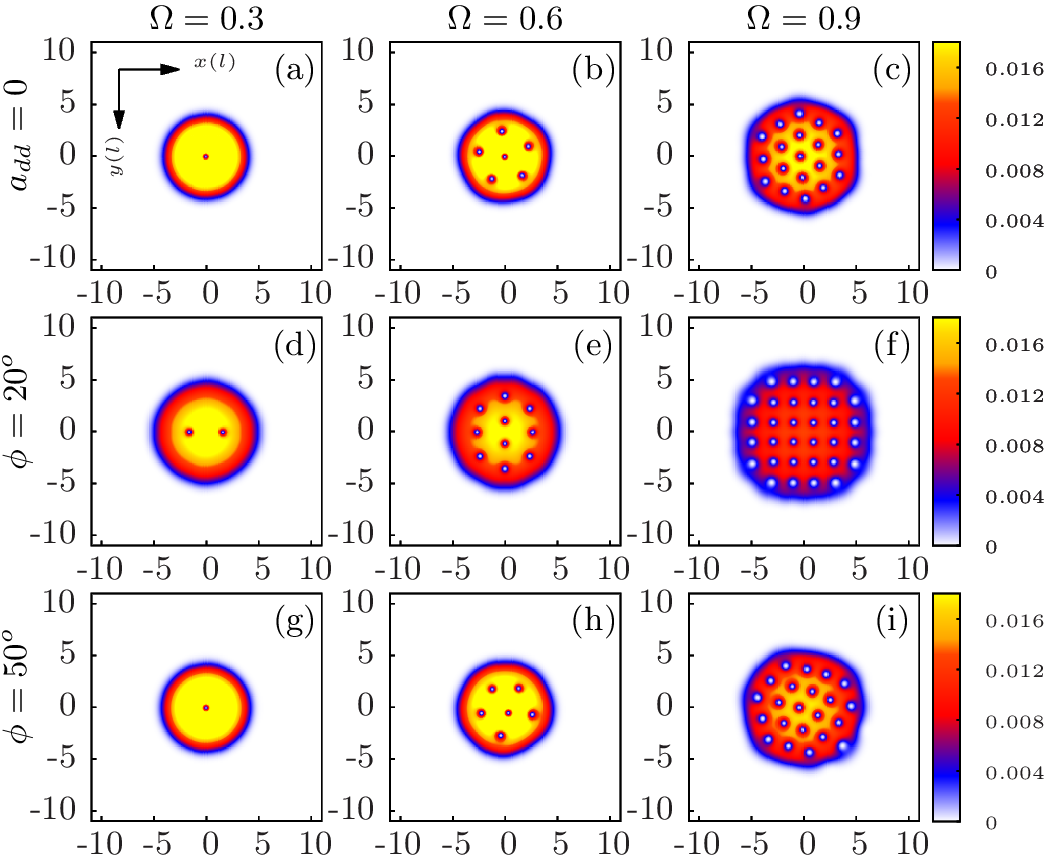}
\caption{Ground-state density profiles showing the evolution of vortex lattice structures in a rotating $^{164}$Dy dipolar Bose-Einstein condensate. The upper panel corresponds to the non-dipolar case ($a_{dd} = 0$), while the middle and lower panels represent dipolar condensates with tilt angles $\phi = 20^\circ$ and $\phi = 50^\circ$, respectively. The columns correspond to increasing rotation frequencies $\Omega = 0.3$, $0.6$, and $0.9$ (from left to right). The panels illustrate the emergence and evolution of vortex structures with increasing $\Omega$, ranging from a central density depletion to few-vortex states, and eventually to well-structured vortex lattices. The length scales are in units of harmonic oscillator length.}
\label{density_w_omega}
\end{figure}

The generation of vortices in rotating condensate has been studied for a variety of systems~\cite{raman_01,xie_18}. At a rotational frequency greater than the critical frequency $\Omega>\Omega_c$, multiple vortices appear and tend to arrange themselves in a lattice structure, called a vortex lattice. This critical value depends on the interplay of competing interaction strengths~\cite{cai_18}. The vortex is a topological defect in which the density drops to zero and the phase winds around the core by $2\pi n$, where $n$ is the charge of the vortex. The arrangement of vortices in a lattice is affected by interatomic interaction. In non-dipolar condensates, the vortices arrange themselves into a triangular Abrikosov triangular lattice~\cite{abo_shaeeer_01}. Such Abikosov lattices were first predicted to occur in quantized flux lines in superconductors~\cite{abrikosov_57}. In the absence of dipolar interaction, we have verified that the vortex lattice acquires triangular structure in rotating condensate. This is evident from the evolution of the condensate density profiles with $\Omega$ at $a_{dd}=0$, as shown in Fig.~\ref{density_w_omega}(a,b,c). The dipolar interaction further introduces richness in the lattice structures as the competition of the DDI and short-range interaction leads to bubble, stripe, and square lattice, depending on many parameters, mainly the ratio of DDI and short-range interaction~\cite{komineas_07,klaus_22}. In particular, the DDI leads to transition of triangular to square vortex lattice by minimizing the ground-state energy for fast rotating condensate~\cite{cooper_05,zhang_05}.

The density profiles of the ground-state as a function of $\Omega$ for two tilt angles below the magic angle are shown in Fig.~\ref{density_w_omega}. Within the repulsive DDI, the dBEC exhibits a structural transition of vortex lattices with orientations. The density profiles at tilt angles $\phi=20^{\degree}$ and $\phi=50^{\degree}$ are shown in Fig.\ref{density_w_omega}(d,e,f) and Fig.\ref{density_w_omega}(g,h,i), respectively. At low rotational frequency $\Omega = 0.3$, the condensate hosts a small number of vortices, exhibiting only shallow density depletion due to centrifugal effects~\cite{zhao_15}. As the rotation frequency increases to $\Omega = 0.6$, discrete quantized vortices nucleate and organize into few-vortex configurations, with the vortex number and arrangement depend on the dipole orientation. Here, vortices in ground-state density distributions tend to arrange in a square lattice, reflecting the modified interaction landscape induced by dipolar anisotropy [Fig.~\ref{density_w_omega}(e)]. In contrast, for the larger tilt angle ($\phi = 50^\circ$), fewer vortices are supported due to the suppressed dipolar repulsion, which effectively reduces the radial extent of the condensate. At a stronger rotation at $\Omega = 0.9$, the vortices organize in an ordered lattice structures. In particular, at $\phi = 20^{\degree}$ a square vortex lattice emerges, whereas the tilt with $\phi = 50^\circ$ favors a triangular vortex lattice. Thus, the interplay between the rotation and dipole orientations shapes the vortex nucleation threshold, lattice geometry, and extent of the condensate [cf. Fig.~\ref{density_w_omega}(f) and Fig.\ref{density_w_omega}(i)].

\begin{figure}[h]
\includegraphics[width=\linewidth]{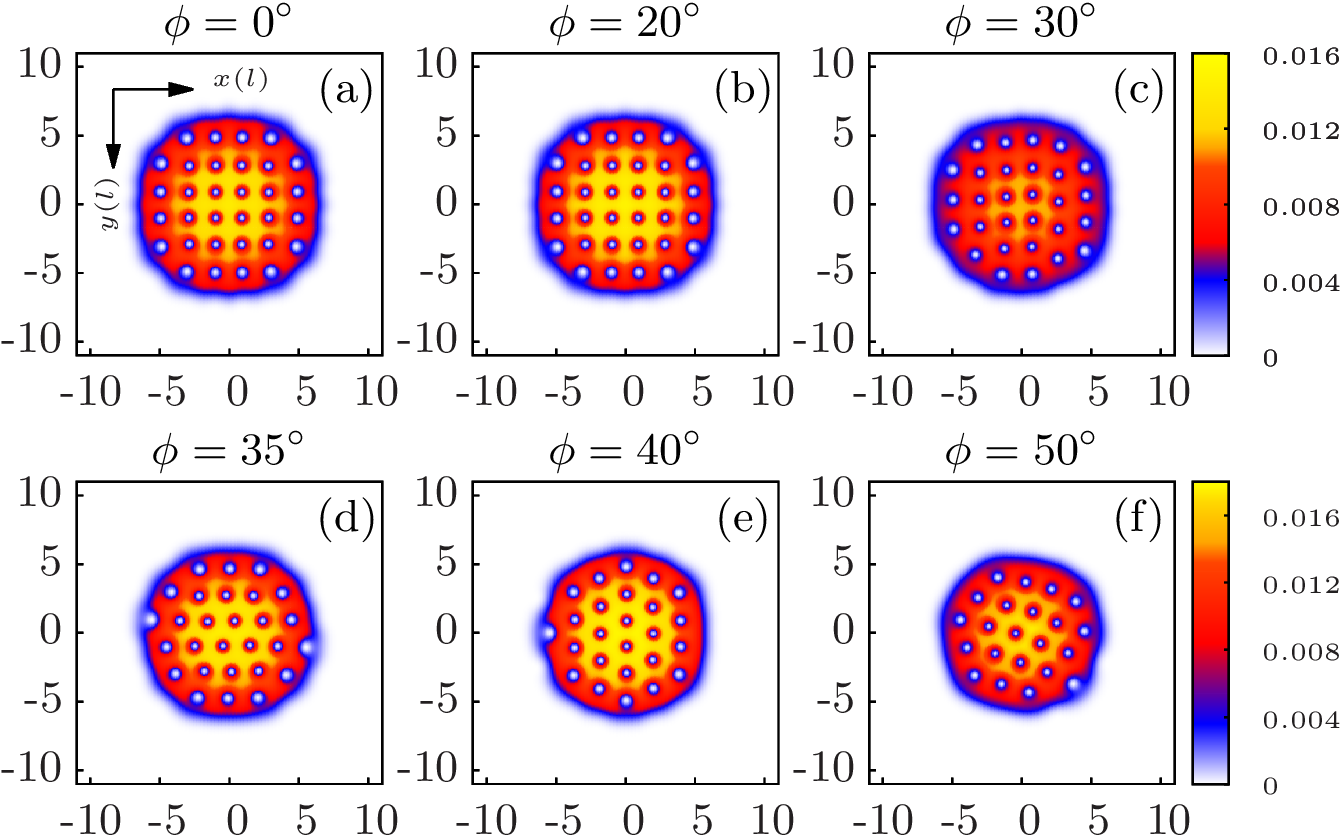}
    \caption{Ground-state density profiles illustrating the evolution of vortex lattices in a rotating $^{164}\mathrm{Dy}$ dipolar Bose-Einstein condensate with tilt angle $\phi$ of dipoles. The rotational frequency is fixed at $\Omega = 0.9$. At small tilt angles, the vortices form a square lattice. With increasing $\phi$, the lattice undergoes a structural transition towards a triangular (Abrikosov-like) configuration. The transition occurs at $\phi_t \approx 35^\circ$, demonstrating the influence of dipolar anisotropy, rotation, and short-range interactions on vortex-lattice stability.}
    \label{density_w_tilt_angle}
\end{figure}

We further analyze the structural transition of vortex lattices with orientation of dipoles for a fast rotating Dy condensate. In Fig.~\ref{density_w_tilt_angle}, we observe that for lower tilt angles, $0^\circ \leqslant \phi \leqslant 35^\circ$ of dipoles, the vortices arrange in  a square lattice. The re-distribution of vortices occurs near the transition angle, which is $\phi_{t}\approx35^{\degree}$. Beyond $\phi_t$, the vortices undergo a structural transition from a square lattice to an Abrikosov triangular lattice. This transition arises from the effective suppression of the dipole-dipole interaction as the tilt angle approaches $\phi_t$. The transition in the lattice structure occurs well below the magic angle, above which the DDI becomes attractive. 
It is worth noting that the configuration of the ground state vortex lattice is strongly influenced by the aspect ratio and relative dipolar interaction strengths. For quasi-2D condensate, a triangular lattice is favoured for $\epsilon_{dd}\lesssim 0.8$ while a square lattice is favoured for $\epsilon_{dd}\gtrsim 0.8$ in the limit when $\Omega$ approaches the transverse confinement frequency~\cite{cai_18,prasad_19}. Our results are consistent with the threshold on $\epsilon_{dd}$ for strongly dipolar Dy condensate. We also explicitly verified that $\phi_{t}$ varies with relative strength $\epsilon_{dd}$; in particular for erbium dBEC the square lattice transition takes place at higher tilt angle as compared to Dy condensate. This is due to comparatively smaller magnetic moment and weaker dipolar strength of erbium condensate. Furthermore, the structure also depends on the dimensionality of the condensate, as recently observed vortex stripes in 3D condensate~\cite{klaus_22}. Thus, the lattice configuration below the magic angle is determined by the interplay of $\epsilon_{dd}$, axial confinement, and orientations of dipoles with respect to the polarization axis.  
\subsection{Role of LHY correction}
 We now turn to discuss the vortex lattice of rotating dBEC when dipoles are tilted above the magic angle. In this case, due to strong dipole-dipole attraction, the condensate collapses for orientations nearly perpendicular to the polarization direction. We incorporate the beyond-mean-field LHY interaction in the rotating frame to stabilize the condensate and vortex lattices therein.  For two-dimensional atomic condensates, the LHY interaction is repulsive in nature, and its inclusion in the system leads to the formation of distinct vortex lattices~\cite{Sabari_26}. We further examine the ground state density distribution of larger number of atoms at different rotational frequencies and orientations.
\begin{figure}[ht]
  \includegraphics[width=\linewidth]{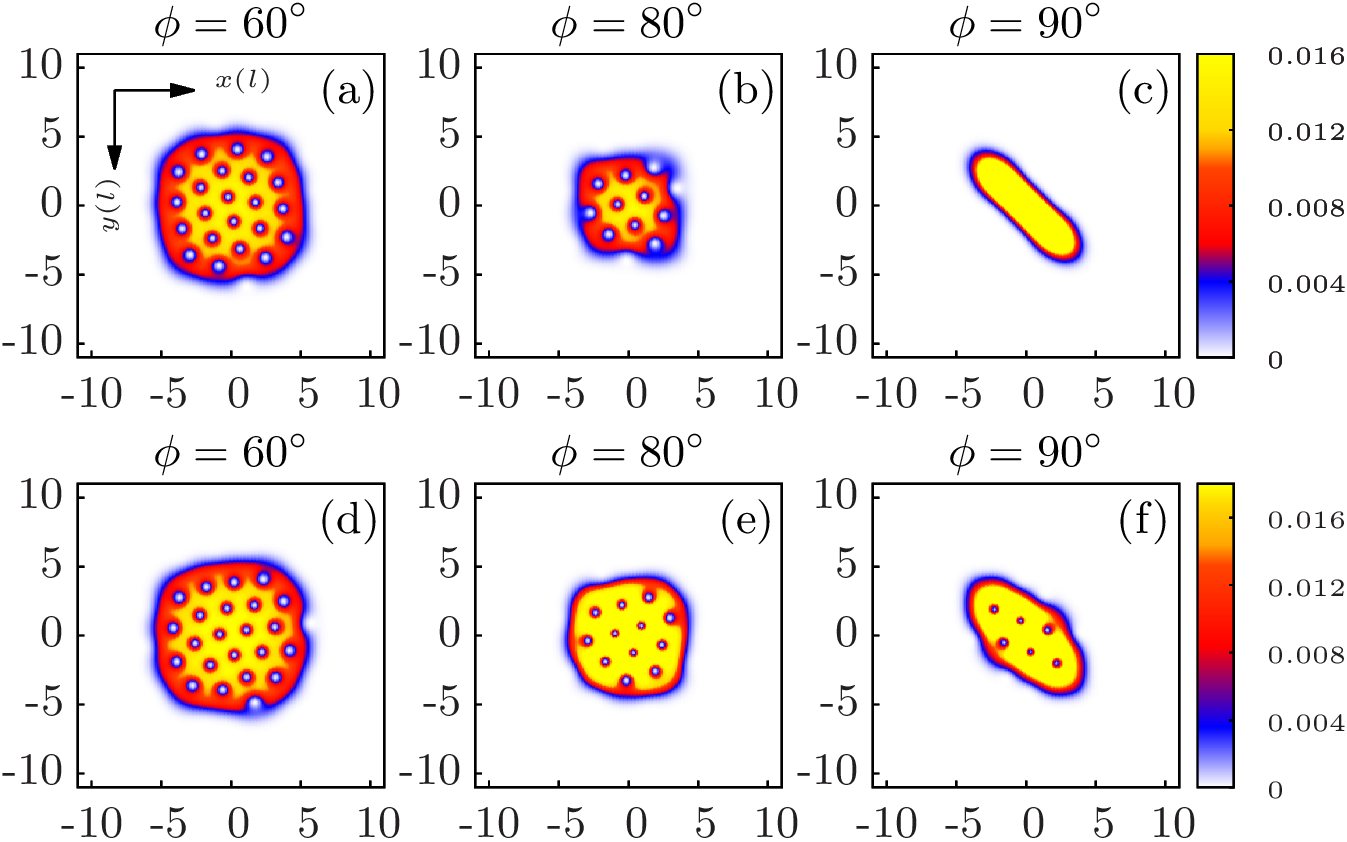}
  \caption{Ground-state density profiles of rotating Dy dipolar condensate with $\Omega=0.95$ at different polarization angles \(\phi=60^{\degree},80^{\degree}, 90^{\degree}\) (a–c) without and (d–f) 
           with the LHY correction. Quantum fluctuations modify the stability landscape, resulting in sustained arrangements of vortices and elongated configurations at perpendicular orientation of dipoles $(\phi=90^{\degree})$. The vertical color-bar indicates normalized density. The length scales are in units of harmonic oscillator length.}
  \label{collapsed_den_pro}
\end{figure}

Fig.~\ref{collapsed_den_pro} presents distinct density profiles of a fast rotating condensate with $\Omega=0.95$ for different tilt angles $\phi$ (with attractive regime of DDI). Here, we reveal the role of LHY correction as $\phi$ is varied for $N=1000$ and $\Omega=0.95$. The upper row of the figure shows the density profiles without the LHY interaction, while the lower row corresponds to the same parameters with the inclusion of the LHY correction. As the attractive interaction dominates with increase in $\phi$, the condensate deforms [Fig.~\ref{collapsed_den_pro}(b)]. The centrifugal effect competes with the radial confinement and eventually the condensate elongates diagonally at $\phi=90^{\degree}$ to minimize its ground state energy [Fig.~\ref{collapsed_den_pro}(c)]. At this tilt angle, the ground-state of the elongated dBEC remains vortex-free. The dynamical geometric squeezing or elongation of interacting condensates have been reported in recent theoretical studies~\cite{sharma_22,chen_25,chen_25a}. In our work, the vortex-free elongation occurs due to the fact that the attraction favours high-density region and energetically prohibits the formation of vortex core (low-density defect) within the condensate. The LHY interaction balances the dominating attractive DDI to stabilize vortices in the condensate. It facilitates the formation of vortices through sustaining the phase winding of the condensate order parameter. At $\phi=60^{\degree}$ and $\phi=80^{\degree}$, the LHY interaction allows to restore the circular symmetry, and more structured vortices, while at $\phi = 90^{\degree}$, the remarkable role of LHY in the nucleation of vortices is evident [cf. Fig.~\ref{collapsed_den_pro}(c) and Fig.~\ref{collapsed_den_pro}(f)]. 

\begin{figure}[h]
\includegraphics[width=\linewidth]{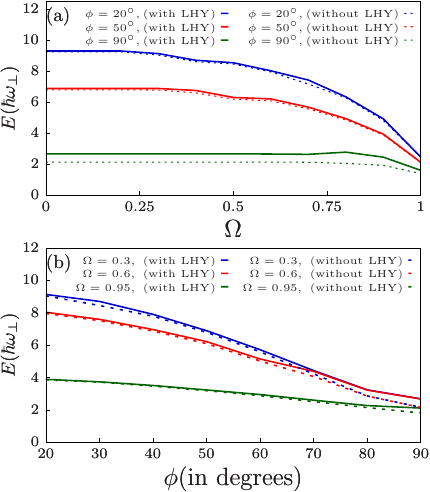}
    \caption{Ground-state energy $E$ as a function of the (a) rotational frequency $(\Omega)$ and (b) dipole's tilt angle $\phi$ for $^{164}\text{Dy}$ dBEC, calculated with and without   
             inclusion of the LHY correction. The LHY correction term increases the total energy, by incorporating the quantum fluctuations in modifying the dependence of anisotropic dipolar interactions and rotation. Here, the ground state energy is in terms of radial quantum energy and tilt angles are in terms of degrees.}
  \label{energy_plot}
\end{figure}

Next, we examine the ground-state energy obtained from Eq.~\eqref{eq:energy} as a function of $\Omega$ and $\phi$ for different sets of parameters. The comparison of energies with respect to the inclusion of repulsive LHY interaction is shown in Fig.~\ref{energy_plot}. In panel (a), the energy is plotted as a function of $\Omega$ for fixed tilt angles $\phi = 20^\circ$, $50^\circ$, and $90^\circ$. These three angles correspond to attractive, nearly zero, and repulsive DDIs. In the absence of the LHY term, the energy decreases with increasing $\Omega$, indicating the enhanced role of rotation in lowering the ground state energy of the system. However, as the tilt angle increases, the DDI becomes increasingly attractive, leading to a significant reduction in energy and eventual instability in $\phi = 90^\circ$. With the inclusion of the LHY interaction, the energy increases compared to without LHY case, reflecting the repulsive quantum fluctuation contributions to the total energy~\cite{Lima_11}. This additional repulsion counteracts the attractive DDI and stabilizes the system even at larger tilt angles, thereby preventing collapse. In panel (b), the energy is shown as a function of the tilt angle $\phi$ for fixed rotational frequencies $\Omega = 0.3, 0.6,$ and $0.95$. Without LHY interaction, the energy decreases sharply as $\phi$ approaches $90^\circ$, signaling the dominance of attractive dipolar interactions and the onset of collapse. In contrast, when the LHY correction is incorporated, the energy variation becomes smoother and remains bounded, demonstrating the stability of the condensate by quantum fluctuations.
\begin{figure}[h]
  \includegraphics[width=\linewidth]{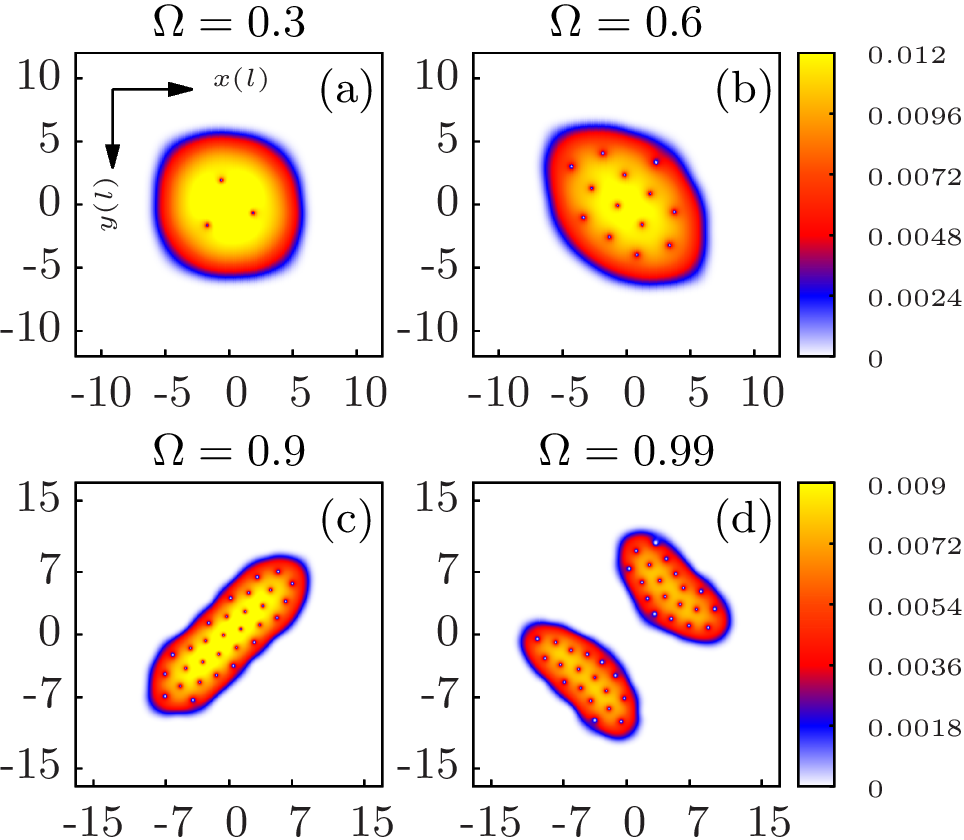}
  \caption{Two-dimensional condensate density profiles of a harmonically trapped $^{164}\mathrm{Dy}$ dBEC, including contact, dipole-dipole, and LHY interactions. The density profiles are 
           presented for a tilt angle $\phi = 90^\circ$ and fixed number of atoms $N = 3\times 10^{4}$. The rotation frequencies are $\Omega = 0.3$ (a), $0.6$ (b), $0.9$ (c), and $0.99$ (d). The spatial extents are expressed in terms of $l$.}
  \label{density_w_natoms}
\end{figure}

We further analyze the case of larger number of dipolar atoms tilted above the magic angle that corresponds to attractive regime. In Fig.~\ref{density_w_natoms}, we present the effect of 
$\Omega$ on the ground state density distributions in the presence of LHY interactions for $N = 3\times 10^{4}$. Considering the perpendicular orientations of the dipoles with $\phi = 90^\circ$, where the DDI is maximally attractive and tends to destabilize the condensate, leading to collapse in the absence of fluctuations. The inclusion of LHY interaction prevents this collapse and gives rise to stable and distinct density configurations. At a lower rotation frequency, $\Omega = 0.3$ [Fig.~\ref{density_w_natoms} (a)], the condensate exhibits a nearly isotropic structure with a few vortices already present. As $\Omega$ is increased to $0.6$ [Fig.~\ref{density_w_natoms}(b)], more vortices nucleate in the condensate which begins to elongate diagonally due to anisotropic DDI. With further increase in rotational frequency to $\Omega = 0.9$ [Fig.~\ref{density_w_natoms}(c)], a condensate becomes significantly elongated, and vortices form a more organized structure along the elongated axis, indicating the strong influence of rotation and dipolar anisotropy. At very high rotation, $\Omega = 0.99$ [Fig.~\ref{density_w_natoms}(d)], centrifugal effects strongly reduce the effective confinement, resulting in the fragmentation of the condensate into two spatially separated high-density lobes, each hosting multiple vortices. The fragments of the condensate possess nearly an equal number of vortices. Thus, the intriguing interplay of dipolar interaction, rotation, and LHY stabilization results in anisotropic density elongation and eventual spatial phase separation. We have explicitly checked that the fragmented condensates remain ground state and stems due to LHY repulsion. Recent previous studies have shown the fragmentation of non-dipolar condensate due to anharmonic (quartic) distortion~\cite{kumar_17,brito_20,dutta_23} while the novel fragmentation phenomenon presented here is due to anisotropic DDI under fast rotation (external trap has circular symmetry with $\gamma = \nu = 1$). It is worth noting that the squeezing of condensate cloud and vortex lattices of fast rotating non-dipolar BEC have been observed recently~\cite{fletcher_21,yao_24}.  
\begin{figure}[h]
  \includegraphics[width=\linewidth]{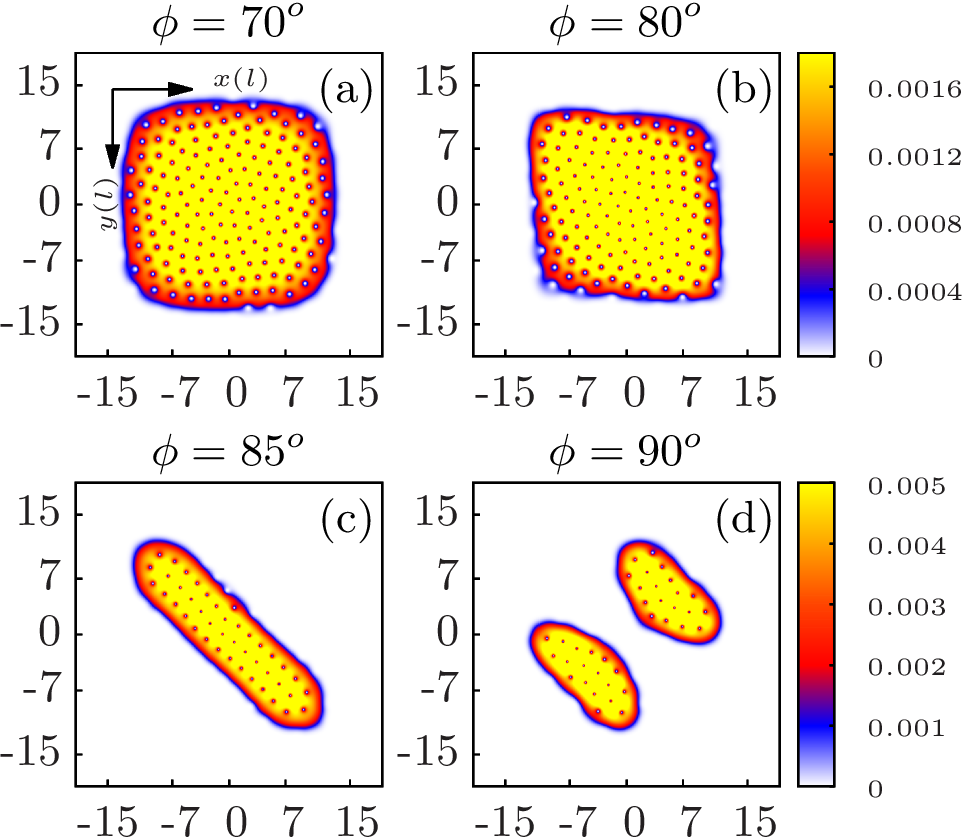}
  \caption{The condensate density profiles of a harmonically confined fast rotating $^{164}\text{Dy}$ dBEC with $\Omega=0.99$ and $N=3\times10^{4}$. The orientation of dipoles is varied in the attractive regime as $\phi=70^\circ$, $80^\circ$, $85^\circ$, and $90^\circ$ [(a)-(d)], respectively. Results are obtained from the extended Gr\"oss-Pitaevskii equation including contact, dipole-dipole, and LHY interactions. Squeezing of the condensate density as well as fragmentation are evident. Here, $x$ and $y$ are in units of $l$.}
  \label{density_w_phi}
\end{figure}

To unveil the role of tilt angle on the breaking-up or fragmentation phenomenon, we examine the change in the density profiles of $^{164}\mathrm{Dy}$ dBEC with LHY interaction at $\Omega=0.99$. The tilt angles are varied within the attractive DDI regime. Fig.~\ref{density_w_phi} shows the distinct role of attractive dipole-dipole interaction and repulsive LHY interaction for $N = 3\times 10^{4}$. In Fig.~\ref{density_w_phi}(a), at $\phi=70^\circ$, the DDI is less attractive and effectively counter-balanced by repulsive LHY interaction hence, the condensate density remains unaffected. However, at $\phi=80^\circ$ the system is more attractive causing the contraction of condensate density [Fig.~\ref{density_w_phi}(b)]. Furthermore, as shown in Fig.~\ref{density_w_phi}(c) and Fig.~\ref{density_w_phi}(d) with $\phi=85^\circ$ and $\phi=90^\circ$, the DDI is highly attractive. At $\phi = 85^\circ$, the condensate undergoes strong anisotropic deformation, forming an diagonally elongated structure with vortices arranged along its long axis. At $\phi = 90^\circ$, where the DDI is maximally attractive, the combined effects of repulsive LHY interaction and the strong centrifugal force overcomes the effective radial confinement of the harmonic trap, leading to the breaking-up of the condensate into two spatially separated fragmented condensates. Thus, the fragmentation of strong dBEC occurs when $\Omega$ approaches the radial confinement frequency and dipoles aligned perpendicular to the polarization axis.  

\subsection{Dynamical nucleation of vortices}
\begin{figure}[h]
  \includegraphics[width=\linewidth]{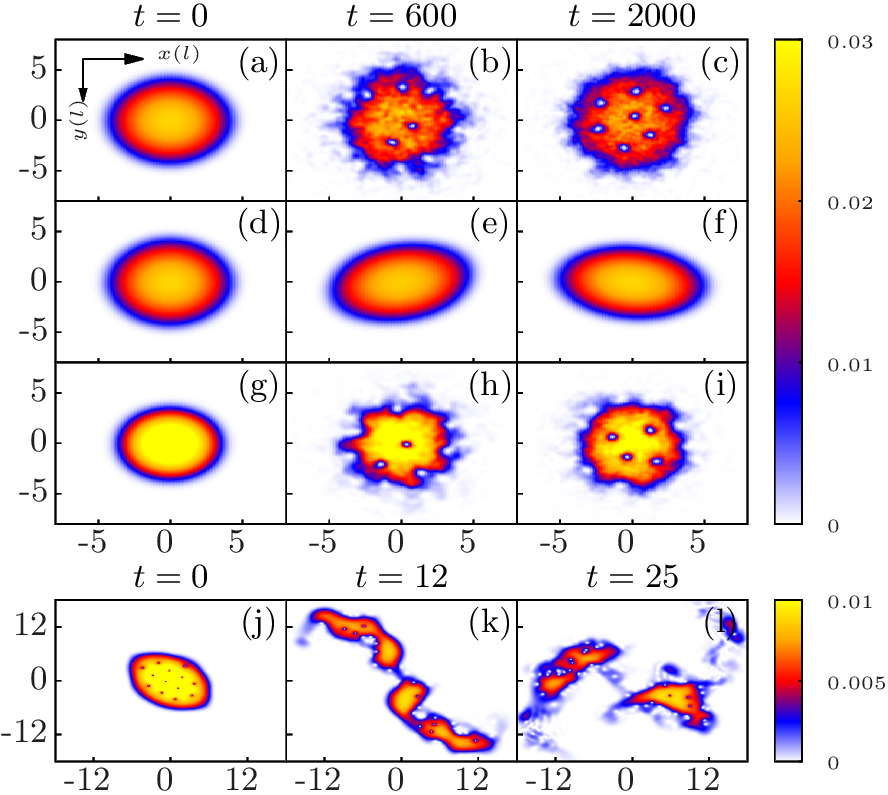}
  \caption{Real-time evolution of the condensate density profile following a rotational frequency quench. The panel (a,b,c) correspond to the evolution with $\Omega$ quench to the value  
           $0.6$ and tilt angle $\phi=40^{\circ}$. With same tilt angle but higher rotational frequency quench $\Omega=0.9$ is shown in panel (d,e,f). The panel (g,h,i) represents the dynamical evolution with $\Omega$ quench at $\Omega = 0.6$ and $\phi=60^{\circ}$ ($\phi>\phi_{m}$). In these cases of sudden quench, the initial (at $t=0$) rotational frequency is $\Omega=0$ and $N = 10^3$. Panels (j–l) show the dynamics for a larger number of atoms ($N = 3\times10^4$) following a quench of the rotational frequency from $\Omega = 0.6$ to $0.99$. The time is presented in scaled units.}
      \label{quench_dynam}
  \end{figure}

We finally discuss the nucleation of vortices in the anisotropic dipolar bosons where the tilting of the dipoles above the magic angle is stabilized by the LHY interaction. The real-time dynamical evolution of the density profile of dBEC following sudden quench of rotational frequency is shown in Fig.~\ref{quench_dynam}. To explore the effect of tilt angle, we first consider the quench dynamics with $N=1000$. We prepare the initial ground state of dBEC at a fixed orientation by solving Eq.~\eqref{eq:GP} using imaginary-time propagation in the absence of angular momentum $(\Omega=0)$. The dBEC is further propagated in real-time by including a finite angular momentum $\Omega\neq 0$ up to $t \sim 2\times 10^{3}$ (in scaled units). A small anisotropy is introduced to facilitate the dynamical formation of vortices~\cite{ripoll_01,tsubota_02,kasamatsu_05}. Here, the nucleation of vortices with time occurs even in the absence of any dissipation in the system. The emergence of vortices becomes energetically favorable for the quench of $\Omega$ above a critical value. 
  
The formation of vortices for the orientation of dipoles with $\phi < \phi_{m}$ is studied. The density profile of the tilted dBEC for $\phi = 40^\circ$ at two rotational frequencies $\Omega=0.6$
and $\Omega=0.9$ is shown in Fig.~\ref{quench_dynam}(a-f). Initially, the condensate density shows no phase singularity at $t=0$. The deformation from the circular symmetry is caused by finite tilt of dipoles. As the system evolves over time, the vortices begin to enter the condensate through surface mode instabilities [Fig.~\ref{quench_dynam}(a,b,c)]. As time progresses, the number of vortices increases; however, the system stabilizes the number of vortices at longer time dynamics. In the present case, the system possesses seven vortices [Fig.~\ref{quench_dynam}(c)], which remains the same with longer time evolution. Similar vortex nucleation dynamics and anisotropy-driven vortex patterns have been reported in earlier theoretical and experimental studies of rotating dipolar condensates~\cite{lahaye_09,prasad_19}.  
When the system is quenched to the $\Omega$ close to trapping frequency [Fig.~\ref{quench_dynam}(d,e,f)], the condensate density evolves with no-vortex state, even in the presence of anisotropy and dissipation. This is attributed to lack of the growth of the surface instability due to a sudden quench. The system thus perform breathing oscillations and stuck in vortex-free state. We have found that the adiabatic quench allows the dynamical appearance of vortices in the condensate (not shown here). Thus, the sudden quench excites the collective excitations instead of topological defects. Furthermore, the quench dynamics of attractive DDI with $\phi=60^\circ$ also nucleates vortices when quenched to $\Omega=0.6$ [Fig.~\ref{quench_dynam}(g,h,i)]. Here, the system gets stabilized with four vortices. Therefore, the nature of the dipolar interaction influences the stability of vortices; in particular the repulsive interaction facilitates the effective transfer of angular momentum within the condensate and leads to larger number of vortices.

As the larger number of atoms $N = 3\times10^{3}$ exhibits fragmentation phenomenon for $\Omega\approx1$ with perpendicular orientation of dipoles, we now consider the quench dynamics. We begin the initial state $\Omega=0.6$ that is quenched close to the trapping frequency with $\phi=90^{\circ}$. Here, the system is stabilized by taking into account the LHY correction. The condensate tends to elongate along the attractive dipolar direction while the centrifugal effect of rotation tries to make it circular. The competition of these two effects leads to squeezing of the condensate and finally breaking in multiple fragments. The fragments carry vortices; however, the deformation due to strong attraction results in less stable vortices. 
\begin{figure}[ht]
  \includegraphics[width=\linewidth]{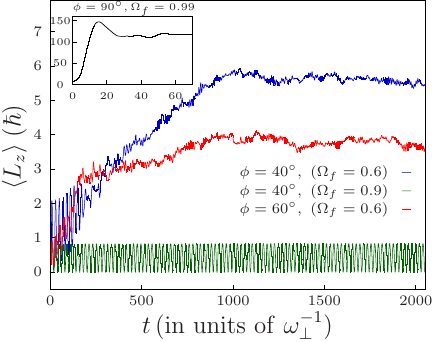}
      \caption{Time evolution of the expectation value of the angular momentum $\langle \hat{L}_z \rangle$ corresponding to Fig.~\ref{quench_dynam}. The change in the angular momentum is associated to the emergence of vortices during sudden quench of rotational frequency. The tilt angle and final values of rotational frequency $\Omega_{f}$ for three cases are shown in the figure. The inset plot shows the angular momentum evolution for large number of atoms corresponding to fragments with vortices. The time is shown in units of inverse of radial trapping frequency.}
      \label{ang_mom}
\end{figure}  

Finally, we investigate the stability of generated vortices by examining the time evolution of the average angular momentum per particle $\langle{\hat{L}_z}\rangle$. It is defined as 
$\langle \hat{L}_{z} \rangle = i\int \psi^{*}(\mathbf{\rho},t) (y\partial_{x} - x\partial_{y}) \psi(\mathbf{\rho},t)d\mathbf{\rho}$ and shown in Fig.~\ref{ang_mom} for various cases of quench dynamics. For the two cases of dynamics, corresponding to the quench $\Omega=0.6$ [Fig.~\ref{quench_dynam}(a,b,c) and Fig.~\ref{quench_dynam}(g,h,i)], the $\langle{\hat{L}_{z}}\rangle$ gradually increases with periodic oscillations and eventually attains a steady value, reflecting the condensate with a stable number of vortices. Moreover, the steady value of $\langle{\hat{L}_{z}}\rangle$ for $\phi = 40^{\circ}$ is larger than that of $\phi = 60^{\circ}$, which is due to comparatively more vortices with repulsive DDI. On the other hand, the $\langle{\hat{L}_{z}}\rangle$ for the rotational quench at $\Omega=0.9$, oscillates near zero and does not acquire a finite steady value. This confirms the vortex-free state of the condensate at strong 
rotational frequency quench. For a large number of atoms i.e. ($N = 3\times10^4$), the $\langle{\hat{L}_{z}}\rangle$ for $\phi = 90^\circ$ and $\Omega$ quench from $0.6$ to $0.99$ [inset of Fig.~\ref{ang_mom}] presents the steady value confirming the vortices in fragments of the system. Hence, the angular momentum confirms the dynamical evolution of dipolar condensate at different rotational quench and dipole orientations.  

\section{Conclusions}
\label{Conclusion}
We have investigated the vortex lattices in the ground state of a harmonically trapped quasi-two-dimensional rotating anisotropic dipolar Bose–Einstein condensate. Our results reveal that the anisotropic nature of dipolar interactions, controlled by the tilt of the magnetic dipoles, plays a crucial role in shaping the structure of vortex lattices. This leads to a transition from a square lattice to a triangular lattice that occurs at approximately $\phi_{t} \approx 35^\circ$ for Dy condensate. When the orientation of dipoles exceeds the magic angle, the dipolar interaction becomes attractive, leading to a notable anisotropic deformation of the condensate. Consequently, the density elongates along a diagonal direction, inhibiting vortex nucleation and ultimately resulting in a vortex-free state; this effect is most significant at $\phi = 90^\circ$ (perpendicular orientation). The inclusion of a beyond mean-field Lee-Huang-Yang correction introduces quantum fluctuations that assist in stabilizing the condensate by promoting vortices in attractive dipolar interaction regimes. With an increased number of atoms, the LHY correction causes the fragmentation of strongly dipolar condensates with vortices as the rotational frequency approaches the radial trapping frequency. Finally, we reveal the dynamical nucleation of vortices through quenching rotational frequency; however, a strong quench induces breathing oscillations in the condensate, and the system remains devoid of vortices over prolonged time dynamics. In light of recent progress in the realization of squeezing dynamics of vortex lattices of non-dipolar condensates~\cite{fletcher_21} and vortices in dipolar condensates~\cite{klaus_22}, our study presents novel quantum phenomena of anisotropic dipolar condensates that could be useful for current dipolar gas experiments. 

\begin{acknowledgments}
 M.A. and K.S. acknowledge support from the Science and Engineering Research Board, Department of Science and Technology, Government of India through Project No. SRG/2023/001569. H.K. acknowledges the financial support from the University Grant Commission (UGC), New Delhi.
\end{acknowledgments}

\bibliographystyle{apsrev4-2} 
\bibliography{references} 
\newpage
  \end{document}